\documentclass[twocolumn,preprintnumbers,amsmath,amssymb,superscriptaddress,longbibliography, prl]{revtex4-1}

\usepackage{graphicx}
\usepackage{bm}
\usepackage{dsfont}
\usepackage[usenames,dvipsnames]{xcolor}
\usepackage{pstricks}
\usepackage[tight]{subfigure}
\usepackage{verbatim}
\usepackage{units}
\usepackage{multirow}
\usepackage{enumitem}
\usepackage{mathrsfs}
\usepackage{leftidx}
\usepackage{xspace}
\usepackage{braket}
\usepackage{bbm}
\usepackage{cancel}
\usepackage{amsfonts,amssymb,amsmath}
\usepackage[normalem]{ulem}

\usepackage{hyperref}
\hypersetup{colorlinks=true,linktoc=all,linkcolor=blue,breaklinks=true,citecolor=blue,urlcolor=blue}
\usepackage[english]{babel}
\newcommand{\Q}{\mathrm{Q}}

\newcommand{\E}{\mathbb{E}}
\newcommand{\Var}{\mathrm{Var}}

\usepackage[most]{tcolorbox}
\definecolor{shadecolor}{RGB}{50,50,50}

\begin{document}

\title{Uncertainty relations for arbitrary currents in coherent transport}

\author{Ludovico Tesser}
	\affiliation{Department of Microtechnology and Nanoscience (MC2), Chalmers University of Technology, S-412 96 G\"oteborg, Sweden}
	
\author{Janine Splettstoesser}
	\affiliation{Department of Microtechnology and Nanoscience (MC2), Chalmers University of Technology, S-412 96 G\"oteborg, Sweden}
	
\date{\today}

\begin{abstract}
    We derive  thermodynamic and  kinetic uncertainty relations valid for arbitrary currents in coherent, strongly coupled,  linear systems out of equilibrium.
    Exploiting properties of the transport statistics, in particular fluctuation theorems, we identify the relevant entropy production and activity that determine the cost of precision at the level of individual scattering events. The resulting bounds include higher-order fluctuations and remain valid far from equilibrium. We illustrate our results in normal and superconducting hybrid structures, and show that their predictiveness and validity range exceeds existing formulations.
\end{abstract}

\maketitle

Transport through nanoscale devices is significantly affected by fluctuations~\cite{Blanter2000Sep, Esposito2009Dec, Landi2024Apr}. Consequently, the precision of currents flowing through them is a crucial quantifier of the performance of such devices~\cite{Campbell2026Jan}.
To characterize the limits on the attainable precision and the cost necessary to achieve high precision, uncertainty relations, such as the thermodynamic uncertainty relation (TUR)~\cite{Barato2015Apr} and the kinetic uncertainty relation (KUR)~\cite{Garrahan2017Mar,DiTerlizzi2018Dec}, have been developed. These trade-off relations are formulated in the generic form
\begin{equation}
    \mathcal{P}^\Q :=\frac{(\dot{Q})^2}{S^\Q} \leq \text{cost},
\end{equation}
where, on the left-hand side, the ratio between an average current ($\dot{Q}$) squared and its zero-frequency noise ($S^\Q$) defines the precision $\mathcal{P}^\Q$ of the generic transported quantity $\Q$.
The right-hand side captures the thermodynamic or kinetic cost necessary to achieve current precision.
In its first formulation, valid for classical Markov processes, the thermodynamic uncertainty relation has as cost the entropy production rate, $\text{cost}\to\dot{\Sigma}/2$ (we set the Boltzmann constant $k_\text{B}=1$), making the attainable precision limited by dissipation~\cite{Barato2015Apr}. Instead, the kinetic uncertainty relation has as cost the dynamical activity, $\text{cost} \to \mathcal{A}_\text{cl}$, which corresponds to the average rate of jumps along stochastic trajectories of the system dynamics~\cite{Garrahan2017Mar,DiTerlizzi2018Dec}. 
The entropy and activity costs can even be unified in the thermokinetic uncertainty relation (TKUR)~\cite{Vo2022Sep}, which reduces to the TUR close to equilibrium and to the KUR far from it, while bridging the two bounds in intermediate regimes.

Subsequent works extended these results to quantum systems described by Lindblad master equations~\cite{Hasegawa2020Jul, Prech2025Jan}.
These extensions are however limited to the weak-coupling regime, beyond which the bounds can be violated~\cite{Brandner2018Mar,Agarwalla2018Oct,Liu2019Jun,Saryal2019Oct,Ehrlich2021Jul, Mahadeviya2026Apr}.
Extending uncertainty relations in strongly-coupled systems is more challenging, as the distinction between system and baths is no longer clear-cut.
Nonetheless, precision limits have recently been established on the currents flowing through systems with negligible particle-particle interactions (or treatable as a mean field)~\cite{Brandner2025Jul, Brandner2025Oct, Palmqvist2025Jul, Palmqvist2025Oct, Blasi2025May}, which are accurately described by scattering theory~\cite{Blanter2000Sep}.
However, these quantum thermodynamic~\cite{Brandner2025Jul} and kinetic~\cite{Blasi2025May} uncertainty relations have been demonstrated \textit{only} for the precision of the particle current $\mathrm{Q}\to\mathrm{N}$.
This limitation becomes particularly relevant in hybrid superconducting structures, where quasiparticle, charge, and energy currents are not equivalent, and where Andreev reflections result in violations of precision limits for the charge current, which were established to hold for the particle current, only~\cite{Ohnmacht2025Dec,Mayo2026Mar, Sobrino2026May}.
This raises the question of how to formulate uncertainty relations for \textit{arbitrary} transported quantities, which remain valid even in hybrid superconducting structures. 

In this Letter, we answer this question by proving precision limits on arbitrary currents described by scattering theory, valid for coherent transport in any type of linear system.
Our approach is based on the properties of the probability distribution of the transport process and enables generalizations of both the thermodynamic \textit{and} the kinetic uncertainty relation. We use fluctuation theorems~\cite{Hasegawa2019Sep, Timpanaro2019Aug, Potts2019Nov} to extend the thermodynamic uncertainty relation to coherent transport, and the probability of changing the transported quantity to define the activity entering the kinetic uncertainty relation~\cite{Tesser2025-thesis,Hasegawa2025Nov,Hegde2026Mar, VanVu2026May}. 
These results remain valid far from equilibrium and beyond weak coupling, and naturally include higher-order contributions of stochastic entropy production and activity that are essential for the validity of the uncertainty relations in coherent conductors.
We demonstrate our results in normal and superconducting hybrid structures, and show that they provide tighter and more general constraints compared to previous formulations.

\textit{Precision limits in coherent transport}---We consider the transport of an arbitrary quantity $\Q$ through a coherent conductor (also called scattering region) coupled to multiple reservoirs~\cite{Blanter2000Sep, Moskalets2011Sep}.
The average current $\dot{Q}$ and zero-frequency noise $S^\Q$ are written as
\begin{equation}\label{eq:current-noise}
    \dot{Q} = \int \frac{dE}{h} \E[q], \quad S^\Q = \int \frac{dE}{h} \Var[q],
\end{equation}
when interactions (or nonlinearities in the system Hamiltonian) are negligible~\cite{Levitov1993charge, Belzig2001Oct}.
Here, the expectation value $\E[\bullet]$ and the variance $\text{Var}[\bullet]$ are taken over the probability distribution $p$ of scattering events at energy $E$, and $q$ is the stochastic variable representing the transported quantity.
As an example,  consider the particle current in reservoir $\alpha$, $\Q\to \mathrm{N}_\alpha$.
The corresponding statistics are determined by the probability $p(n)$ of transferring $n\in\mathbb{Z}$ particles in reservoir $\alpha$, and $q(n)=n$ is the stochastic number of particles added to reservoir $\alpha$.
To keep the discussion general, in the following both $p$ and $q$ are evaluated on events $\omega$ in a sample space $\Omega$.
Concrete examples are provided below for (quasi)particle, charge, and energy transport between normal conductors and through a normal-superconducting junction.

To derive the thermodynamic uncertainty relation, we first consider an involution on the sample space, i.e. a map $\bullet^\ddagger:\Omega\to\Omega$ such that $(\omega^\ddagger)^\ddagger = \omega$ for all $\omega$.
We focus on quantities $q(\omega)$ that are \textit{antisymmetric} with respect to the involution, namely
\begin{equation}\label{eq:TUR-q-antisymmetry}
    q(\omega) \stackrel{!}{=} -q(\omega^\ddagger).
\end{equation}
In the example of particle current, where the number $n$ of transferred particles fully characterizes an event $\omega$, a suitable involution fulfilling this condition would be $n^\ddagger=-n$ for the amount of transferred particles.
This example is a special case of the time-reversal involution, in which a stochastic trajectory is mapped to the reversed trajectory of the time-reversed system. The time-reversal involution is particularly appealing in time-reversal symmetric systems, where the probability of a trajectory and its reversed are linked by the thermodynamic entropy produced in the trajectory~\cite{Landi2021Sep,Strasberg2022Book}. 

Without assuming that the system is time-reversal symmetric, we define the forward-backward stochastic entropy $\tilde{\sigma}$~\cite{Tesser2025-thesis,VanVu2026May} and the symmetrized probability distribution $p_+$ as
\begin{subequations}\label{eq:TUR-prob-definitions}
    \begin{align}
    \Tilde{\sigma}(\omega) &:= \log \frac{p(\omega)}{p(\omega^\ddagger)},\label{eq:sigma-definition}\\
    p_+(\omega) &:= \frac{p(\omega)+p(\omega^\ddagger)}{2}.
    \end{align}
\end{subequations}
The forward-backward stochastic entropy is antisymmetric by construction, $\Tilde{\sigma}(\omega) = -\Tilde{\sigma}(\omega^\ddagger)$, and the symmetrized probability $p_+(\omega)$ is a valid probability distribution on the sample space $\Omega$. We denote with $\E_+[\bullet]$ the expectation values taken with respect to $p_+$.
Note that, for time-reversal symmetric systems, the forward-backward entropy production coincides with the actual thermodynamic entropy production $\sigma$.

By definition, $\tilde{\sigma}$ fulfills a fluctuation theorem [Eq.~\eqref{eq:sigma-definition}], so we can follow the strategy of Refs.~\cite{Hasegawa2019Sep, Timpanaro2019Aug, Potts2019Nov} to derive a thermodynamic uncertainty relation.
Using Eq.~\eqref{eq:TUR-prob-definitions}, we write the first and second moment of $q$ as expectations on $p_+(\omega)$,
\begin{equation}
    \E[q] = \E_+\left[q\tanh\frac{\Tilde{\sigma}}{2}\right],\quad \E[q^2] = \E_+[q^2].
\end{equation}
We then apply the Cauchy-Schwarz inequality on the symmetrized distribution and map the resulting expressions back to the original probability space, namely
\begin{subequations}\label{eq:TUR-moments}
\begin{align}
    \E_+\left[q\tanh\frac{\Tilde{\sigma}}{2}\right]^2 &\leq \E_+[q^2]\E_+\left[\left(\tanh\frac{\Tilde{\sigma}}{2}\right)^2\right], \\
    \E[q]^2 &\leq  \E[q^2]\E\left[\tanh\frac{\Tilde{\sigma}}{2}\right],
\end{align}
\end{subequations}
where we used the antisymmetry of $\tanh(\tilde{\sigma}/2)$.
In terms of the first two cumulants, it then reads
\begin{equation}\label{eq:TUR}
    \Var[q] \geq \E[q]^2 \frac{1-\E\left[\tanh\frac{\Tilde{\sigma}}{2}\right]}{\E\left[\tanh\frac{\Tilde{\sigma}}{2}\right]}.
\end{equation}
We now combine Eq.~\eqref{eq:current-noise} with Cauchy-Schwarz inequality (this time on integrable functions) proving
\begin{equation}\label{eq:scatteringTUR}
     \frac{(\dot{Q})^2}{S^\Q}\leq \int\frac{dE}{h}\frac{\E\left[\tanh\frac{\Tilde{\sigma}}{2}\right]}{1-\E\left[\tanh\frac{\Tilde{\sigma}}{2}\right]}=:\mathcal{C}.
\end{equation}
This TUR for coherent transport is the first central result of the paper.
When the forward-backward entropy is small $\tilde{\sigma} \ll 1$, we approximate Eq.~\eqref{eq:scatteringTUR} as
\begin{equation}\label{eq:scatteringTUR-expansion}
     \frac{(\dot{Q})^2}{S^\Q}\leq \int\frac{dE}{h}\E\left[\frac{\Tilde{\sigma}}{2}\right] + \mathcal{O}(\tilde{\sigma}^2) \approx \frac{1}{2}\dot{\tilde{\Sigma}},
\end{equation}
where $\dot{\tilde{\Sigma}}$ is the average forward-backward entropy production rate.
Thus, close to equilibrium, we recover the classical formulation of the thermodynamic uncertainty relation, which is expected since the linear response approximation holds~\cite{Macieszczak2018Sep, Kheradsoud2019Aug}.
Far from equilibrium, expanding the right-hand side of Eq.~\eqref{eq:scatteringTUR} shows that higher moments of the entropy production have significant (positive) contributions to the cost, allowing coherent conductors to achieve higher precision than the classical limit.

Note that it is possible to replace the higher-order statistics of the entropy production with higher powers of the average entropy production by employing Jensen's inequality on the cost of precision, as done in Ref.~\cite{Zhang2019Oct}. At the price of a looser bound, one can therefore formulate the cost in terms of the average entropy production $\E[\tilde{\sigma}]$, only. 

To derive the kinetic uncertainty relation, we instead introduce the stochastic variable
\begin{equation}\label{eq:a-definition}
    a(\omega):= 1-\delta_{q(\omega),0}
\end{equation}
counting the events in which a non-zero amount of $q$ is transferred.
Clearly, $\E[q] = \E[aq]$ by construction, and by employing Cauchy-Schwarz inequality we find
\begin{equation}\label{eq:KUR-moments}
    \E[q]^2 = \E[aq]^2 \leq \E[a]\E[q^2]
\end{equation}
where we used $a^2=a$. In terms of the cumulants of $q$, Eq.~\eqref{eq:KUR-moments} reads
\begin{equation}\label{eq:KUR}
    \Var[q] \geq \E[q]^2 \frac{1-\E[a]}{\E[a]}
\end{equation}
where $\E[a]$ is the probability that the transferred quantity changes, and it plays the role of an activity~\cite{Tesser2025-thesis,VanVu2026May}.
Applying Eq.~\eqref{eq:KUR} and Cauchy-Schwarz inequality (on integrable functions) we find the KUR for coherent transport
\begin{equation}\label{eq:scatteringKUR}
    \frac{(\dot{Q})^2}{S^\Q}\leq \int\frac{dE}{h}\frac{\E[a]}{1-\E[a]}=:\mathcal{A},
\end{equation}
which is the second central result of the paper.
We stress that the stochastic activity $a$ is defined in terms of the specific quantity $q$ of interest. However, in a similar fashion to Eq.~\eqref{eq:TUR-q-antisymmetry}, one can consider the activity for a \textit{class} of quantities. A natural example provided by the involution $\bullet^\ddagger$ is the class of quantities $q$ fulfilling $q(\omega)=0$ for all $\omega = \omega^\ddagger$.
However, even more restrictive classes of quantities may be of interest. Consider, for instance, a multi-terminal setup, in which the probability distribution describing particle transfer between the $r$ reservoirs reads $p(n_1,\cdots, n_r)$.
For the global reversal involution $(n_1,\cdots, n_r)^\ddagger=(-n_1,\cdots, -n_r)$ only the null event $(0,\cdots,0)$ contributes to $\E[1-a]$. If, however, one is interested only in the current flowing into a particular terminal, say the first one, a local reversal involution $(n_1, n_2,\cdots, n_r)^\ddagger=(-n_1, n_2,\cdots, n_r)$ suffices. This increases the number of events contributing to $\E[1-a]$ and yields a tighter constraint on the precision.

Similarly to Eq.~\eqref{eq:scatteringTUR-expansion}, the kinetic uncertainty relation Eq.~\eqref{eq:scatteringKUR} has an appealing expansion when the probability of changing $\Q$ is small, $\E[a]\ll1$. This happens in the limit where the reservoirs are weakly coupled to the scattering region, such that the transmission probability from one reservoir to another becomes small. Then, the right-hand side of Eq.~\eqref{eq:scatteringKUR} is expanded as
\begin{equation}\label{eq:scatteringKUR-expansion}
    \frac{(\dot{Q})^2}{S^\Q}\leq \int\frac{dE}{h}{\E[a]} + \mathcal{O}(\E[a]^2)\approx \mathcal{A}_\text{cl},
\end{equation}
where the rate $\mathcal{A}_\text{cl}$ quantifies the number of events transferring a non-zero amount of quantity $\Q$ per unit time.
Similarly to the TUR, as one moves away from the weak-coupling limit, higher powers of $\E[a]$ will have a significant (positive) contribution to the precision limit. This means that strongly coupled systems can achieve higher precision than in the weakly-coupled limit, as was also noted in Refs.~\cite{Palmqvist2025Jul, Palmqvist2025Oct, Blasi2025May}, albeit in different formulations.


We now merge the two approaches by considering both antisymmetric quantities $q$ [Eq.~\eqref{eq:TUR-q-antisymmetry}], and the stochastic activity $a$ [Eq.~\eqref{eq:a-definition}].
Combining Eqs.~(\ref{eq:TUR-moments}, \ref{eq:KUR-moments}), we have
\begin{equation}
    \left(\E_+\left[aq\tanh\frac{\tilde{\sigma}}{2}\right]\right)^2 \leq \E_+[q^2]\E_+\left[a\,\tanh\frac{\tilde{\sigma}}{2}\right],
\end{equation}
which leads to the energy-resolved thermokinetic uncertainty relation (TKUR)
\begin{equation}\label{eq:TKUR}
    \Var[q] \geq \E[q]^2 \frac{1-\E\left[a\,\tanh\frac{\tilde{\sigma}}{2}\right]}{\E\left[a\,\tanh\frac{\tilde{\sigma}}{2}\right]},
\end{equation}
and finally to the TKUR for coherent transport
\begin{equation}\label{eq:scatteringTKUR}
    \frac{(\dot{Q})^2}{S^\Q} \leq \int \frac{dE}{h} \frac{\E\left[a\,\tanh\frac{\tilde{\sigma}}{2}\right]}{1-\E\left[a\,\tanh\frac{\tilde{\sigma}}{2}\right]}.
\end{equation}
How much tighter Eq.~\eqref{eq:scatteringTKUR} is compared to the TUR of Eq.~\eqref{eq:scatteringTUR} depends on the class of quantities with respect to which $a$ is defined.
Indeed, if $a(\omega)=0$ only on the events $\omega=\omega^\ddagger$ for which $\tilde{\sigma}(\omega)=0$, then $\E[a\,\tanh(\tilde{\sigma}/{2})] =\E[\tanh(\tilde{\sigma}/{2})] $, and Eq.~\eqref{eq:scatteringTKUR} reduces to Eq.~\eqref{eq:scatteringTUR}.
This shows that (i) a more selective definition of activity $a$ leads to a tighter TKUR~\eqref{eq:scatteringTKUR}, and (ii) the TUR~\eqref{eq:scatteringTUR} already includes activity-like contributions from events with $\omega=\omega^\ddagger$.
For simplicity, in the following examples we only consider cases in which Eq.~\eqref{eq:scatteringTKUR} reduces to Eq.~\eqref{eq:scatteringTUR}.

\textit{Examples}---We now illustrate our main results [Eqs.~(\ref{eq:scatteringTUR}, \ref{eq:scatteringKUR})] for coherent fermionic transport between normal conductors (NN setup) and in a normal-superconducting junction (NS setup).
We take the involution $\bullet^\ddagger$ to exchange the initial and final states of the scattering process and focus on time-reversal symmetric setups. In this case, the forward-backward entropy production $\tilde{\sigma}$ coincides with the thermodynamic entropy production, making the thermodynamic constraint of Eq.~\eqref{eq:scatteringTUR} physically insightful.

We compare the TUR~\eqref{eq:scatteringTUR} with the particle-current bound derived in Ref.~\cite{Brandner2025Jul}, which reads
\begin{equation}\label{eq:quantum_TUR}
    \frac{(\dot{N}_\alpha)^2}{S^{\text{N}_\alpha}} \leq  \dot{N}_\alpha \sinh\left(\frac12 \frac{\dot{\Sigma}}{\dot{N}_\alpha}\right)=: \mathcal{C}_\text{qu}
\end{equation}
and which holds for fermionic, time-reversal symmetric systems. This relation reduces to the classical TUR close to equilibrium, where $\dot{\Sigma}\to 0$, and, remarkably, is only slightly modified for time-reversal symmetry breaking~\cite{Brandner2025Jul}.
However, unlike the classical TUR, Eq.~\eqref{eq:quantum_TUR} was only proven for particle-current precision.
Violations were shown for the charge-current precision $\mathcal{P}^\text{C}$ in the presence of superconductivity, when the quasiparticle current $\dot{N}_\alpha$ in the cost is replaced by the charge current normalized by the electron charge $\dot{C}_\alpha/e$~\cite{Ohnmacht2025Dec, Mayo2026Mar, Sobrino2026May}.

Furthermore, we compare the KUR~\eqref{eq:scatteringKUR} with the particle-current bound derived in Ref.~\cite{Blasi2025May}, which reads
\begin{equation}\label{eq:quantum_KUR}
    \frac{(\dot{N}_\alpha)^2}{S^{\text{N}_\alpha}} \leq \frac{(\mathcal{A}_\text{cross})^2}{\mathcal{A}_\text{cross} - \mathcal{A}_\text{sh}}=:\mathcal{A}_\text{qu},
\end{equation}
where the terms $\mathcal{A}_\text{cross/sh}$ emerge from different decompositions of the tunnel coupling fluctuations, and are given by
\begin{subequations}
    \begin{align}
        \mathcal{A}_\text{cross} &:=\!\int\!\! \frac{dE}{h}\sum_{\beta\neq\alpha} \tau_{\alpha\beta}[f_\alpha(1-f_\beta)+f_\beta(1-f_\alpha)],\label{eq:Across}\\
        \mathcal{A}_\text{sh} &:=\!\int\!\! \frac{dE}{h}\sum_{\beta\neq\alpha} \tau_{\alpha\beta}[f_\alpha-f_\beta]^2.
    \end{align}
\end{subequations}
Here, $f_\alpha(E)=(1+\exp[(E-\mu_\alpha)/T_\alpha])^{-1}$ is the Fermi distribution in bath $\alpha$ (at temperature $T_\alpha$ and chemical potential $\mu_\alpha$), and $\tau_{\alpha\beta}$ is the transmission probability from $\beta$ to $\alpha$.
While the definition of activity in Ref.~\cite{Blasi2025May} has appealing connections to the quantum Fisher information~\cite{Palmqvist2026, Palmqvist2026Jul}, the precision limit in Eq.~\eqref{eq:quantum_KUR} was only proven for the precision of the particle current.

\begin{figure}
    \centering
    \includegraphics[page=1]{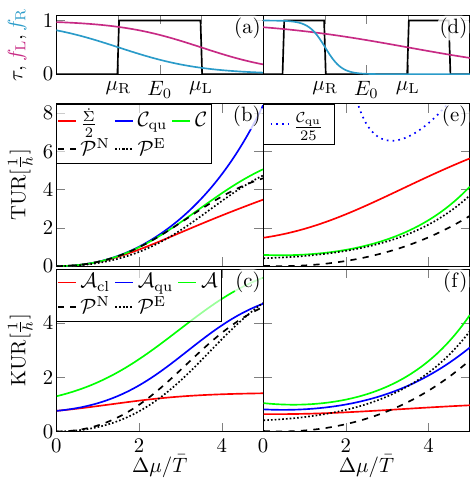}
    \caption{Precision limits in NN setup. (a,d) transmission $\tau$ and occupations $f_\text{L,R}$ as functions of energy at $\Delta\mu=2\bar{T}$. (b,c,e,f) Particle ($\mathcal{P}^\text{N}$) and energy ($\mathcal{P}^\text{E}$) current precision, with (b,e) thermodynamic costs from Eqs.~(\ref{eq:scatteringTUR-expansion}, \ref{eq:quantum_TUR}, \ref{eq:scatteringTUR}) and (c,f) kinetic costs from Eqs.~(\ref{eq:scatteringKUR-expansion}, \ref{eq:quantum_KUR}, \ref{eq:scatteringKUR}) as functions of $\Delta\mu$. In (a-c) $\Delta T=0$ and $\tau(E) = \Theta[\frac{w^2}{4}-(E-E_0)^2]$ is a boxcar. In (d-f) $\Delta T = 0.8\bar{T}$ and $\tau(E) = \Theta[{w^2}-(E-E_0)^2]-\Theta[\frac{w^2}{4}-(E-E_0)^2]$ is a double boxcar. In all plots we fix $\mu_\text{R}=0, E_0=\bar{T},$ and $w=2\bar{T}$.}
    \label{fig:NN}
\end{figure}

First, we consider an NN setup made of a single-channel two-terminal conductor with transmission probability $\tau(E)$.
Particles incoming from the left/right lead have occupations $f_\text{L/R}(E)$ given by Fermi distributions with average temperature $\bar{T} = (T_\text{L}+T_\text{R})/2$, temperature bias $\Delta T = T_\text{L}-T_\text{R}$, and chemical potential bias $\Delta\mu = \mu_\text{L} - \mu_\text{R}$. 
Transport is fully determined by the probability $p(n_\text{L})$ of adding $n_\text{L}$ ($=-n_\text{R}$ by particle number conservation) particles in the left lead after a scattering event, and is given by (suppressing the energy dependence $E$)~\cite{Levitov1993charge,Tesser2026Apr}
\begin{equation}
    p(-1) = \tau f_\text{L}(1-f_\text{R}),\quad  p(1)= \tau (1-f_\text{L})f_\text{R},
\end{equation}
and $p(0) = 1-p(1)-p(-1)$, since we are considering fermions.
Using Eq.~\eqref{eq:sigma-definition}, the forward-backward entropy production reads
\begin{equation}
    \tilde{\sigma}(1) = \log\frac{(1-f_\text{L})f_\text{R}}{f_\text{L}(1-f_\text{R})} = \frac{E-\mu_\text{L}}{T_\text{L}} -\frac{E-\mu_\text{R}}{T_\text{R}}
\end{equation}
conciding with the thermodynamic entropy production, as expected. The expectation values entering the TUR~\eqref{eq:scatteringTUR} and the KUR~\eqref{eq:scatteringKUR} are
\begin{subequations}
\begin{align}
    \E\left[\tanh\frac{\tilde{\sigma}}{2}\right] &= \tanh\left(\frac{\tilde{\sigma}(1)}{2}\right) \tau(f_\text{R}-f_\text{L}),\\
    \E[a] &= \tau [f_\text{L}(1-f_\text{R})+f_\text{R}(1-f_\text{L})],
\end{align}
\end{subequations}
respectively. The first differs from the energy-resolved entropy production rate by the tanh, while the second coincides with the integrand in Eq.~\eqref{eq:Across} and corresponds to the single-particle contribution to the particle-current noise (sometimes also called classical noise contribution).

In Fig.~\ref{fig:NN}, we illustrate the relation between the precision limits derived here and those found previously for coherent transport in linear systems~\cite{Blasi2025May,Brandner2025Jul}.
The transmission $\tau$ is a boxcar [Fig.~\ref{fig:NN}(a-c)] or a double boxcar [Fig.~\ref{fig:NN}(d-f)] since sharp features in the transmission probability are instrumental for violations of the classical TUR~\cite{Ehrlich2021Jul, Brandner2025Jul}.
For the boxcar transmission, these violations happen at large $\Delta \mu$ [Fig.~\ref{fig:NN}(b)], where both particle ($\mathcal{P}^\text{N}$) and energy ($\mathcal{P}^\text{E}$) current precision exceed $\dot{\Sigma}/2$. The inequality is recovered when considering the thermodynamic cost of Eqs.~(\ref{eq:quantum_TUR}, \ref{eq:scatteringTUR}).
The difference between the two costs becomes evident far from equilibrium [Fig.~\ref{fig:NN}(e), where the temperature bias $\Delta T$ is comparable with the average temperature $\bar{T}$], where Eq.~\eqref{eq:scatteringTUR} provides a much tighter constraint compared to Eq.~\eqref{eq:quantum_TUR}.
This is expected because $\mathcal{C}$ (i) contains significant contributions from higher moments of the entropy production and (ii) already contains activity-like contributions, as discussed below Eq.~\eqref{eq:scatteringTKUR}.

The classical KUR is also violated at large $\Delta\mu$ [Fig.~\ref{fig:NN}(c,f)], making the corrections of Eqs.~(\ref{eq:quantum_KUR}, \ref{eq:scatteringKUR}) necessary.
While the kinetic cost $\mathcal{A}_\text{qu}$ [Eq.~\eqref{eq:quantum_KUR}] provides a tighter constraint for the particle-current precision $\mathcal{P}^\text{N}$ compared to Eq.~\eqref{eq:scatteringKUR}, it does in general not limit the precision of the energy current ($\mathcal{P}^\text{E}$), as seen for the example of the double boxcar transmission [Fig.~\ref{fig:NN}(f)], where transport happens in separate energy windows~\cite{Eriksson2021Sep}. 
This shows that Eq.~\eqref{eq:quantum_KUR} does not provide a general extension of the classical KUR to the coherent transport regime, and we consequently no longer use it as a comparison.

\begin{figure}
    \centering
    \includegraphics[page=2]{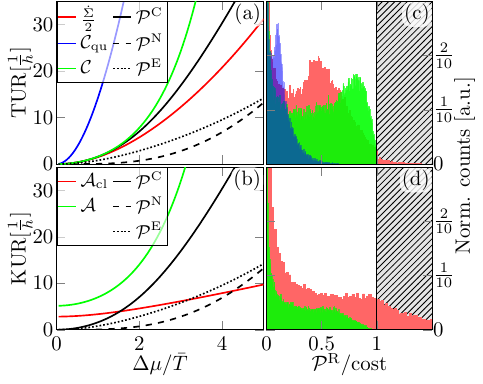}
    \caption{Precision limits in NS setup. (a,b) Charge ($\mathcal{P}^\text{C}$), quasiparticles ($\mathcal{P}^\text{N}$), energy ($\mathcal{P}^\text{E}$) current precision with  (a) thermodynamic costs from Eqs.~(\ref{eq:scatteringTUR-expansion}, \ref{eq:quantum_TUR}, \ref{eq:scatteringTUR}) and (b) kinetic costs from Eqs.~(\ref{eq:scatteringKUR-expansion}, \ref{eq:scatteringKUR}), as functions of $\Delta \mu$. (c,d) Sampling of precision $\mathcal{P}^\text{R}$ of random energy-independent observable for $\tau\in[0,1], \Delta\mu\in[0,5\bar{T}]$ divided by (c) thermodynamic costs of Eqs.~(\ref{eq:scatteringTUR-expansion}, \ref{eq:quantum_TUR}, \ref{eq:scatteringTUR}) or (d) kinetic costs of Eqs.~(\ref{eq:scatteringKUR-expansion}, \ref{eq:scatteringKUR}). In (a,b) $\tau=1$. In (c,d) the hatched region highlights violations of classical TUR and KUR. In all plots $\bar{T}=\Delta,\Delta T=0$.}
    \label{fig:NS}
\end{figure}

We now turn to the example of the NS setup, where the right lead is superconducting with gap $\Delta$. We describe the transport statistics using BTK theory~\cite{Blonder1982Apr} with a uniform (normal-state) transmission probability $\tau$.
In contrast to the NN setup, the quasiparticles describing transport in the NS setup do not all carry the same charge, so the particle and charge current can differ significantly.
This difference was exploited in Refs.~\cite{Ohnmacht2025Dec,Mayo2026Mar, Sobrino2026May} to show violations of Eq.~\eqref{eq:quantum_TUR} when the particle current and its noise are replaced, \textit{both} in the precision and in the cost $\mathcal{C}_\text{qu}$, by the charge current $\dot{C}_\alpha/e$ (normalized by the electron charge $e$) and its noise.
Here, we instead keep the quasiparticle current in the thermodynamic cost of Eq.~\eqref{eq:quantum_TUR} because a (stochastic) entropy change only happens when there is a (stochastic) quasiparticle transfer. This is not the case for the charge current, where Andreev reflections transfer charge without changing entropy, and hence allow the charge-current precision $\mathcal{P}^\text{C}$ to violate the classical TUR [Fig.~\ref{fig:NS}(a)].
When the quasiparticle current is used in the cost  $\mathcal{C}_\text{qu}$, the charge-current precision $\mathcal{P}^\text{C}$ does \textit{not} violate Eq.~\eqref{eq:quantum_TUR}.
Similarly to the NN setup, the cost $\mathcal{C}$ of Eq.~\eqref{eq:scatteringTUR} provides a tighter constraint on the precision.
This is corroborated by Fig.~\ref{fig:NS}(c), where we sample the ratio between the precision and the thermodynamic costs of Eqs.~(\ref{eq:scatteringTUR-expansion}, \ref{eq:quantum_TUR}, \ref{eq:scatteringTUR}) for random energy-independent local observables ($\mathcal{P}^\text{R}$) and for randomly chosen $\tau, \Delta\mu$.
While the classical TUR shows violations, Eqs.~(\ref{eq:quantum_TUR}, \ref{eq:scatteringTUR}) do not, and the cost $\mathcal{C}_\text{qu}$ yields a narrower distribution around zero compared to $\mathcal{C}$.

The classical KUR is also violated at large $\Delta\mu$ [Fig.~\ref{fig:NS}(b)] by the precision of both charge ($\mathcal{P}^\text{C}$), quasiparticle ($\mathcal{P}^\text{N}$), and energy ($\mathcal{P}^\text{E}$) currents. Notably, Andreev reflections---which contribute to charge transport only---make $\mathcal{P}^\text{C}$ overcome the limit set by the classical KUR at lower $\Delta\mu$.
The inequality is recovered when considering the kinetic cost of Eq.~\eqref{eq:scatteringKUR}.
Figure~\ref{fig:NS}(d) shows that violations of the classical KUR happen more frequently compared to the classical TUR [Fig.~\ref{fig:NS}(c)], but, as in the NN setup, the thermodynamic uncertainty relation [Eq.~\eqref{eq:scatteringTUR}] gives a tighter constraint than the KUR [Eq.~\eqref{eq:scatteringKUR}].

\textit{Conclusions}---In this Letter we have established thermodynamic, kinetic, and thermokinetic uncertainty relations for coherent, strongly-coupled transport.
Unlike prior formulations, our approach based on the transport probability distribution is not limited to the particle-current precision alone, but holds for arbitrary transport quantities.
Our results clearly show that, far from equilibrium or at strong coupling, the thermodynamic and kinetic costs contain higher-order contributions, making precisions violate the classical bounds.
The generality of the approach makes it suitable not only to transport between normal conductors, but also in the presence of superconducting leads, where the classical uncertainty relations can be violated without the need of a transmission function rapidly varying in energy.

In this work, we focused on time-reversal symmetric setups, where the forward-backward entropy $\tilde{\sigma}$ coincides with the thermodynamic entropy production. When time-reversal symmetry is broken, for instance by magnetic fields, the precision can exceed the limits set by the thermodynamic entropy production~\cite{Macieszczak2018Sep,Potts2019Nov,Taddei2023Sep, Barker2025Nov}, while bound~\eqref{eq:TUR} remains valid.
Clarifying the physical meaning of the corresponding forward-backward entropy in this case is an important direction for future work.

Furthermore, Eqs.~(\ref{eq:scatteringTUR}, \ref{eq:scatteringKUR}, \ref{eq:scatteringTKUR}) require the average current and noise to take on the form of Eq.~\eqref{eq:current-noise}, valid for noninteracting systems.
The fundamental question of how to extend the probability-based framework discussed here to the interacting case is still open.
The recent progress in Ref.~\cite{Palmqvist2026Jul} on the KUR for arbitrary quantum systems suggests that the Fisher information is key to such an extension.

\textit{Acknowledgements}---We thank Didrik Palmqvist, Kay Brandner, and Mark Mitchison for the useful discussions, and Elsa Danielsson and Sofia Sevitz for providing useful comments on the draft.
We acknowledge financial support from the Knut and Alice Wallenberg foundation through the fellowship program and from the European Research Council (ERC) under the European Union’s Horizon Europe research and innovation program (101088169/NanoRecycle).

\bibliography{refs.bib}

\end{document}